\documentclass[sigplan,10pt]{acmart}

\AtBeginDocument{%
  }

\usepackage{amsmath}
\usepackage{booktabs}
\usepackage{array}
\usepackage{graphicx}

\newcommand{\pf}{\textsc{PrefixFlow}}

\begin{document}
\title{PrefixFlow: Training-Time KV Caching for Schedule-Level Prefix Reuse in LLM RL Training}

\settopmatter{authorsperrow=3}
\makeatletter
\newcommand{\affilmark}[1]{%
  \expandafter\gdef\csname typeset@author\the\num@authors\endcsname##1{##1\textsuperscript{#1}}}
\def\institution#1{\global\@ACM@instpresenttrue\ignorespaces}
\def\city#1{\global\@ACM@citypresenttrue\ignorespaces}
\def\country#1{\global\@ACM@countrypresenttrue\ignorespaces}
\newcommand{\@affiliationlegend}{%
  \mbox{\textsuperscript{1}Hong Kong University of Science and Technology}\quad
  \mbox{\textsuperscript{2}Tencent}\quad
  \mbox{\textsuperscript{3}Shanghai University of Finance and Economics}\quad
  \mbox{\textsuperscript{4}The University of Hong Kong}}
\patchcmd{\@mkauthors@iii}
  {\and\par\bigskip}
  {\and\par\medskip{\@affiliationfont\@affiliationlegend\par}\bigskip}
  {}{\errmessage{authors.tex: could not add the affiliation list to the acmart author block}}
\makeatother

\author{Pengbo Li}\affilmark{1}
\affiliation{%
  \institution{Hong Kong University of Science and Technology}
  \city{Hong Kong}
  \country{China}}
\email{plibt@connect.ust.hk}

\author{Feiyuan Zhang}\affilmark{1}
\affiliation{%
  \institution{Hong Kong University of Science and Technology}
  \city{Hong Kong}
  \country{China}}
\email{fzhangax@connect.ust.hk}

\author{Guangming Sheng}\affilmark{2}
\affiliation{%
  \institution{Tencent}
  \city{Shenzhen}
  \country{China}}
\email{petersheng@tencent.com}

\author{Guangxin He}\affilmark{1}
\affiliation{%
  \institution{Hong Kong University of Science and Technology}
  \city{Hong Kong}
  \country{China}}
\email{gheaf@connect.ust.hk}

\author{Di Chai}\affilmark{3}
\affiliation{%
  \institution{Shanghai University of Finance and Economics}
  \city{Shanghai}
  \country{China}}
\email{chaidi@mail.sufe.edu.cn}

\author{Ziniu Li}\affilmark{2}
\affiliation{%
  \institution{Tencent}
  \city{Shenzhen}
  \country{China}}
\email{ziniuli@tencent.com}

\author{Taiqiang Wu}\affilmark{4}
\affiliation{%
  \institution{The University of Hong Kong}
  \city{Hong Kong}
  \country{China}}
\email{takiwu@connect.hku.hk}

\author{Han Tian}\affilmark{1}
\affiliation{%
  \institution{Hong Kong University of Science and Technology}
  \city{Hong Kong}
  \country{China}}
\email{htianab@connect.ust.hk}

\author{Wenyu Mao}\affilmark{2}
\affiliation{%
  \institution{Tencent}
  \city{Shenzhen}
  \country{China}}
\email{wenyumao2@gmail.com}

\author{Binhang Yuan}\affilmark{1}
\affiliation{%
  \institution{Hong Kong University of Science and Technology}
  \city{Hong Kong}
  \country{China}}
\email{biyuan@ust.hk}

\author{Kai Chen}\affilmark{1}
\affiliation{%
  \institution{Hong Kong University of Science and Technology}
  \city{Hong Kong}
  \country{China}}
\email{kaichen@cse.ust.hk}

\renewcommand{\shortauthors}{Pengbo Li et al.}

\ccsdesc[500]{Computing methodologies~Distributed computing methodologies}
\ccsdesc[500]{Computing methodologies~Machine learning}
\keywords{large language models, reinforcement learning, prefix reuse, KV caching}

\begin{abstract}
In reinforcement learning (RL) post-training, one prompt fans out into \(N\) trajectories, yet standard trainers replay the shared prefix's forward and backward pass \(N\) times; in long-context retrieval-augmented generation (RAG) and multimodal workloads, this repetition dominates training compute.  Existing designs reuse the prefix within one computation graph (a packed sequence, fused kernel, or tree traversal); under the trainer's memory limits and parallelism, their compute savings do not reliably become shorter steps.

We present \pf{}, which takes the prefix out of that graph and into the training schedule through a \emph{training-time KV cache}: Phase A runs one prefix forward and caches its keys and values (\(K/V\)); Phase B trains suffixes as ordinary microbatches that read only \(K/V\) and write back only their gradients (\(gK/gV\)); Phase C runs one delayed prefix backward on the accumulated \(gK/gV\).  Memory and parallelism then become manageable: dormant prefix activations are offloaded during Phase~B to free high-bandwidth memory (HBM) for suffixes, and context, pipeline, and data parallelism each schedule the phases through a local rule while tensor and expert parallelism need no change, so saved compute becomes shorter steps under any microbatch split.

On dense models, and on mixture-of-experts (MoE) models under Rollout Routing Replay, \pf{} matches full-sequence training within bfloat16 (BF16) tolerance.  Against an identically configured baseline, it speeds up end-to-end verl RL training on a long-prefix RAG workload by \(1.90{\times}\) ($6.1 \times$ on the actor update).  In controlled sweeps, actor-side speedup grows with prefix ratio and group size, reaching \(4.4{\times}\) at an 83\% prefix ratio, while suffix-dominated splits gain little; offloading dormant prefix activations lowers Phase-B peak HBM by up to \(59\%\) and extends trainable sequences by \(1.66{\times}\).
\end{abstract}
\copyrightyear{2026}
\acmYear{2026}
\setcopyright{cc}
\setcctype{by}
\acmConference[SIGOPS ATC 2026]{The 2026 ACM SIGOPS Annual Technical Conference}{November 15-18, 2026}{Hong Kong, China}
\acmBooktitle{The 2026 ACM SIGOPS Annual Technical Conference (SIGOPS ATC 2026), November 15-18, 2026, Hong Kong, China}
\acmDOI{10.1145/3828523.3857100}
\acmISBN{979-8-4007-2818-1/2026/11}
\pagestyle{empty}
\maketitle

\section{Introduction}

Reinforcement learning (RL) has become a central stage in post-training large language models (LLMs), from mathematical reasoning and instruction following to long-context retrieval, multimodal reasoning, and agentic interaction~\cite{instructgpt,ppo,deepseekmath,deepseekr1,tulu3,openrlhf,verl}.  A common trainer-side shape is a rollout group: one prompt produces multiple trajectories, and the policy update consumes sequences of the form
\[
    [P \Vert S^{(1)}], [P \Vert S^{(2)}], \ldots, [P \Vert S^{(N)}],
\]
where the prefix \(P\) is identical across the group and each suffix \(S^{(i)}\) is a different rollout.  Standard trainers process these sequences as ordinary training samples or microbatches, so the shared prefix is evaluated repeatedly: prefix forward and prefix backward are both replayed once per trajectory.

This repeated-prefix computation is a real bottleneck in the workloads we target.  Prefixes in long-context RL can contain retrieved passages, image or video tokens, chart or diagram content, tool schemas, system instructions, and the user query.  High-prefix workloads are common but not universal: across fixed-context retrieval-augmented generation (RAG)~\cite{rag}, multimodal, and long-context question-answering (QA) RL tasks, median prefix ratios range from roughly 76\% to 99\% (Table~\ref{tab:motivation-workloads}; Section~\ref{sec:background} details the measurement methodology).  \pf{} targets workloads whose shared prefixes are long enough to dominate compute and whose rollout groups are large enough to amortize one-time prefix work; suffix-dominated workloads, such as math/code reasoning and software-engineering (SWE) agent traces, offer little reuse.

Recent training-side designs place one copy of the shared prefix and all \(N\) suffixes in a single computation graph---a packed sequence, a fused attention kernel, or a tree traversal with relayed state~\cite{prefixgrouper,dpoprefixsharing,dualkv,arealdta,treetrainingv5}---so that backpropagation through that graph keeps the prefix gradients correct.  Binding a group to one graph, however, collides with the rest of the trainer: prefix activations stay resident until the graph finishes backward, context parallelism (CP) cannot balance prefix and suffix tokens that now do different amounts of work, and pipeline parallelism (PP) needs many microbatches to keep its pipeline full, whereas one graph per group yields few.  The compute that reuse removes therefore does not reliably become shorter steps or free memory.

\pf{} instead decouples the prefix and suffix graphs and gives the prefix phases of its own: Phase A runs the shared prefix forward once and caches per-layer keys and values (\(K/V\)); Phase B trains suffixes as ordinary microbatches that read the cached \(K/V\) and accumulate prefix-side gradients \(gK/gV\); Phase C injects the accumulated gradients and runs one delayed prefix backward.  This works because the only interface between a suffix graph and the prefix graph is the \(K/V\) the suffix reads and the \(gK/gV\) it writes back, which together form a \emph{training-time key-value (KV) cache}: an inference KV cache plus a backward half.  We call this approach \emph{schedule-level prefix reuse}: reuse becomes a property of the training schedule instead of one graph, kernel, or traversal.  Concurrent work DualKV~\cite{dualkv} independently identified the same interface but applies it inside a fused kernel, so its reuse stays within one kernel invocation (Section~\ref{sec:related}).  With separate phases, each conflict above can be resolved: dormant prefix activations can leave high-bandwidth memory (HBM) during Phase~B, CP can balance the prefix and suffix phases separately, and PP can keep as many microbatches as it needs, since reuse no longer depends on their number.

\begin{table}[t]
\centering
\caption{Motivating workload statistics. Lengths are median token counts; \(K=1000\).}
\label{tab:motivation-workloads}
\scriptsize
\setlength{\tabcolsep}{2pt}
\begin{tabular}{@{}p{0.22\columnwidth}p{0.25\columnwidth}p{0.15\columnwidth}p{0.15\columnwidth}p{0.14\columnwidth}@{}}
\toprule
Regime & Examples & Prefix & Suffix & Ratio \\
\midrule
\multicolumn{5}{@{}l}{\textit{Prefix-heavy targets}} \\
Fixed RAG & MuSiQue; 2Wiki & 0.9--2.5K & 0.2--0.4K & 80.4--87.6\% \\
Long-context RL/QA & \begin{tabular}[t]{@{}l@{}}LoongRL;\\LongTraceRL\end{tabular} & 18--121K & \(\sim\)0.5--2.0K & \(\sim\)90--99\% \\
Video RLVR & \begin{tabular}[t]{@{}l@{}}Video-MME;\\Video-R1\end{tabular} & 0.8--4.9K & 0.09--0.14K & 85.9--98.0\% \\
Chart QA & ChartQA; AI2D & 0.35--0.64K & 0.13K & 75.8--83.3\% \\
\midrule
\multicolumn{5}{@{}l}{\textit{Low-prefix controls}} \\
Math/code & Open-R1 & 0.1--0.9K & 5.0--12.0K & 2.3--8.0\% \\
SWE agents & OpenHands & 1.7K & 12.8K & 12.8\% \\
\bottomrule
\end{tabular}
\end{table}

Neither the split nor its benefits come for free: three challenges remain (Section~\ref{sec:challenges}), and our contributions address them in turn:

\noindent\textbullet\ \textbf{Schedule-level prefix reuse} (Section~\ref{sec:design}).  C1 is correctness without a shared graph.  We show that nothing but \(K/V\) and \(gK/gV\) crosses between the graphs and that prefix backward obeys \emph{prefix-gradient superposition}: for a fixed prefix forward trace, it is linear in the adjoints arriving at the \(K/V\) interface, so running it once on the accumulated \(gK/gV\) equals, over real arithmetic, running it per suffix and summing the results.  Shared-prefix forward and backward thus run once per rollout group, independent of suffix count, microbatch split, or whether suffixes are organized by padding or packing.

\noindent\textbullet\ \textbf{A system-level training-time KV cache} (Sections~\ref{sec:design-summary} and~\ref{sec:impl-phase-b-hbm}).  C2 is the memory lifecycle of dormant prefix state.  The KV cache doubles as a lifecycle boundary for memory management: \(K/V\) and \(gK/gV\) stay resident, while dormant prefix activations are offloaded during Phase~B and prefetched for Phase~C.  The released HBM goes to longer suffixes, larger suffix microbatches, or fuller suffix activation retention.

\noindent\textbullet\ \textbf{A parallel-aware training implementation} (Sections~\ref{sec:impl-tp}--\ref{sec:impl-dp-placement}).  C3 is the distributed cost of separate phases, such as per-layer \(K/V\) gathers under CP and phase barriers under PP.  Tensor and expert parallelism (TP, EP) need no change; CP, PP, and data parallelism (DP) each reduce to a local balancing, scheduling, or placement rule rather than a new communication protocol, and we state the costs these rules introduce.  Mixture-of-experts (MoE) models compose with \pf{} under Rollout Routing Replay~\cite{rolloutroutingreplay}.

Our evaluation validates these claims on dense and MoE models against a baseline with the same loss, precision, and parallel configuration.  \pf{} matches full-sequence optimizer updates across TP/CP/PP/EP configurations, including an MoE model under Rollout Routing Replay, and preserves a 100-step real RL actor-update trace over an 8.19B-parameter checkpoint within bfloat16 (BF16) tolerance.  Our representative result is a \(1.90{\times}\) end-to-end speedup over 200 closed-loop verl RL steps on a long-prefix RAG workload (\(6.1{\times}\) on the actor update), with divergence from the baseline no larger than a seed change.  Controlled sweeps show the range: actor-side speedup grows with prefix ratio and group size, up to \(4.395{\times}\) at an 83\% prefix ratio, and stays small for suffix-dominated splits; offloading dormant prefix activations reduces Phase-B peak HBM by up to \(59.1\%\) and extends the sequence-capacity frontier by \(1.657{\times}\).

\section{Background and Motivation}
\label{sec:background}

\subsection{GRPO Training Anatomy}

\begin{figure}[t]
    \centering
    \includegraphics[width=.84\columnwidth]{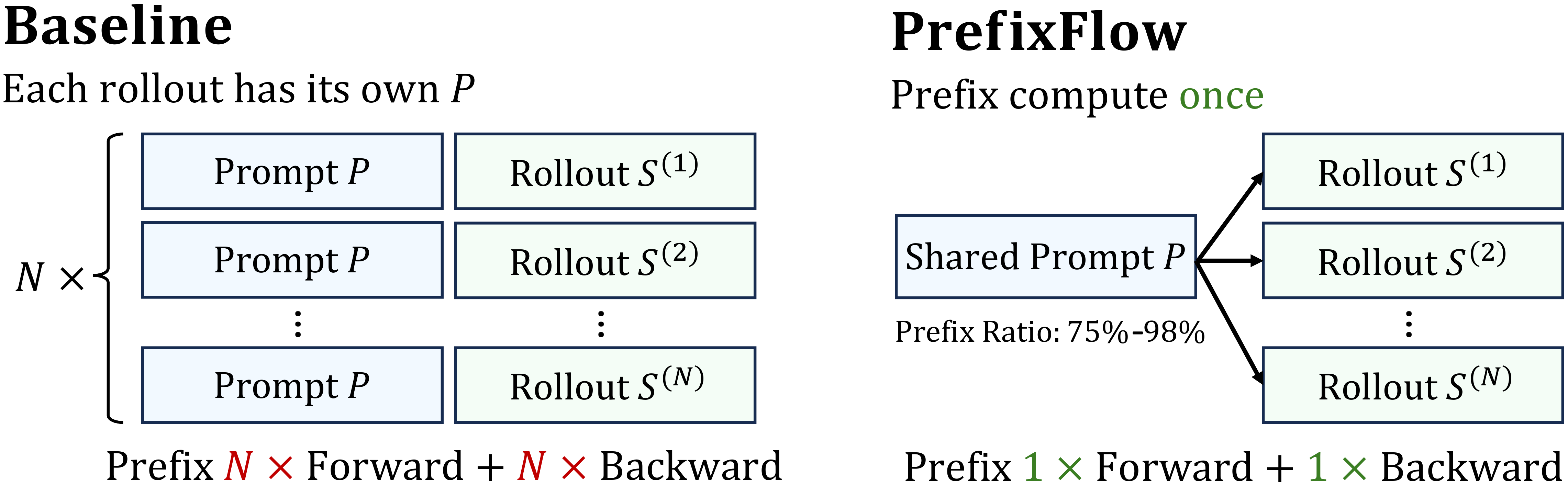}
    \caption{A standard trainer materializes the shared prompt \(P\) in every trajectory, so prefix forward and backward are each executed \(N\) times per rollout group.  \pf{} computes the shared prefix once---one prefix forward and one prefix backward---while every rollout suffix still trains as an ordinary microbatch.}
    \Description{Side-by-side comparison.  Left: a baseline rollout group in which each of the N sequences carries its own copy of prompt P, annotated with prefix N-times forward plus N-times backward.  Right: PrefixFlow, where one shared prompt P with prefix ratio 75 to 98 percent fans out to N rollout suffixes, annotated with prefix one-time forward plus one-time backward.}
    \label{fig:flat-prefix-reuse}
\end{figure}

In group relative policy optimization (GRPO)~\cite{deepseekmath}, a prompt produces a group of \(N\) rollouts.  A trainer computes log probabilities and losses over the generated suffix tokens and then applies one optimizer step to the policy.  The relevant shape for this paper is \emph{flat prefix sharing}: \([P \Vert S^{(1)}]\), \([P \Vert S^{(2)}]\), \ldots, \([P \Vert S^{(N)}]\), a single shared prompt-level prefix with no intermediate branching.
The prefix \(P\) is byte- and token-identical within the group.  Each suffix \(S^{(i)}\) is different and carries its own loss \(L^{(i)}\).  A standard trainer executes \(N\) full forward/backward passes over \([P\Vert S^{(i)}]\), so all prefix-side transformer work is repeated \(N\) times.  Figure~\ref{fig:flat-prefix-reuse} contrasts this with \pf{}: the shared prefix is computed once---one prefix forward and one prefix backward---while every suffix still trains as an ordinary microbatch.

Two details make training-side prefix reuse different from inference caching.  First, the loss may be suffix-only, but gradients still flow to prefix-side parameters through attention.  Suffix queries attend to prefix keys and values~\cite{transformer}, and the resulting loss produces nonzero gradients for prefix \(K/V\) projections and earlier prefix computation.  The shared prefix is therefore not read-only during training.  Second, the trainer may organize suffixes with ordinary batch-plus-padding or with sequence packing~\cite{packing}.  A prefix-reuse system should preserve the baseline masking and position IDs in either layout.

Long shared prefixes are common in the workloads we target (Table~\ref{tab:motivation-workloads}).  We define the prefix ratio as \( |P|/(|P|+|S|) \): the prefix is the input fixed before generation, the suffix is the generated response or released continuation, and per-sample ratios are aggregated into percentiles, combining frozen-rollout measurements with token statistics from public datasets.  Fixed-context RAG~\cite{musique,twowiki}, video RL with verifiable rewards (RLVR)~\cite{videomme,videor1}, and chart/diagram QA~\cite{chartqa,ai2d} rollouts have median prefix ratios from 75.8\% to 98.0\%, while public long-context RL/QA datasets such as LoongRL~\cite{loongrl} and LongTraceRL~\cite{longtracerl} expose shared prompts or contexts at 18K--121K median prefix tokens.  Section~\ref{sec:eval-claim3} places these workloads on a measured prefix-ratio sweep.  The same measurements show the boundary of the opportunity: math/code reasoning~\cite{openr1} and full SWE agent traces~\cite{openhands} are often suffix-dominant because the reasoning chain, tool outputs, or environment history grows much longer than the initial prompt, so \pf{} is not a universal accelerator for RL workloads.

\subsection{Related and Concurrent Prefix-Reuse Designs}
\label{sec:related}

Training-side prefix reuse is easiest to see in one packed graph: place one copy of \(P\) next to the suffixes and mask attention so that suffixes do not attend to each other; autograd then aggregates prefix-side gradients because the prefix appears once in the graph.  Prefix Grouper~\cite{prefixgrouper} and DPO Prefix Sharing~\cite{dpoprefixsharing} follow this pattern with a packed graph or a prompt-pair microbatch, and both are single-machine designs.  DualKV~\cite{dualkv}, concurrent work\footnote{DualKV first appeared on arXiv on May 14, 2026, less than four weeks before our submission.} that independently identified the same \(K/V\)--\(gK/gV\) interface, moves the shared prompt into a custom FlashAttention~\cite{flashattention,flashattention2} kernel whose backward accumulates shared \(gK/gV\) with 32-bit floating-point (FP32) atomics; its reuse lives inside each kernel invocation, and the trainer must co-locate a prompt's responses in one microbatch.

Tree-structured systems relay prefix state across a traversal: AREAL-DTA~\cite{arealdta} keeps the current path's KV and activation state on a depth-first search (DFS) stack, and Tree Training v5~\cite{treetrainingv5}, another concurrent update, serializes tree trajectories in DFS order with differentiable partition gateways.  The relay binds prefix state to traversal order: every leaf's backward in AREAL-DTA re-traverses the shared prefix, so prefix backward is repeated rather than collapsed, and Tree Training's partitioning is a memory-forced split inside one tree sample.

Inference serving systems such as vLLM and SGLang also make prefix reuse central through paged or radix KV-cache management~\cite{vllm,sglang}, and shared-prefix attention kernels avoid re-reading a common prompt for every decoding sequence~\cite{hydragen,chunkattention,flashinfer}.  Both are forward-only: reused \(K/V\) accelerates decoding but does not carry training gradients.  \pf{} imports the KV-cache viewpoint into training and adds the backward counterpart, namely the \(gK/gV\) gradients accumulated across suffix microbatches.  Table~\ref{tab:related-comparison} compares these designs along three axes, which reappear as the challenges C1--C3 of \pf{}'s design (Section~\ref{sec:challenges}).

\begin{table}[t]
\centering
\caption{Training-side prefix-reuse designs along three co-design axes: prefix--suffix coupling (C1), prefix-activation memory lifecycle (C2), and composition with training parallelism (C3).  DualKV and Tree Training v5 are concurrent work; the table compares design properties, not measured performance.  FSDP: fully sharded data parallelism; SP: Ulysses sequence parallelism.}
\label{tab:related-comparison}
\scriptsize
\setlength{\tabcolsep}{2pt}
\begin{tabular}{@{}>{\raggedright\arraybackslash}p{0.23\columnwidth}>{\raggedright\arraybackslash}p{0.22\columnwidth}>{\raggedright\arraybackslash}p{0.21\columnwidth}>{\raggedright\arraybackslash}p{0.22\columnwidth}@{}}
\toprule
System & C1: prefix--suffix coupling & C2: prefix-HBM lifecycle & C3: parallelism \\
\midrule
Prefix Grouper & fused in packed graph & resident with graph & single-machine \\
DPO Prefix Sharing & fused in packed pair & resident with graph & single-machine \\
AREAL-DTA & DFS-bound; bwd per leaf & DFS stack liveness & DP dispatch only \\
Tree Training v5 & tree-serialized + gateways & live gateway state & Megatron/MoE runs only \\
DualKV & kernel-fused, co-located & dedup only, no lifecycle & FSDP/SP/MoE runs only \\
\textbf{\pf{}} & \textbf{decoupled 3-phase schedule} & \textbf{offload + HBM reinvest} & \textbf{TP/EP/CP/PP/DP rules} \\
\bottomrule
\end{tabular}
\end{table}

\subsection{Why Schedule-Level Prefix Reuse Matters}
\label{sec:why-schedule}

These designs share one structural property: reuse lives inside a single invocation---a packed graph, a kernel launch, or a traversal---that holds a group's prefix and suffix computation together.  Inside it the redundant prefix compute is gone, but the rest of the trainer still schedules the invocation as one unit, so the saved compute does not reliably become a shorter step or free memory.  

Under \emph{context parallelism}, prefix tokens only provide their \(K/V\) after reuse while suffix tokens still run the full per-token transformer flow, and a policy that balances both kinds of tokens together leaves suffix-heavy ranks on the critical path.  

Under \emph{pipeline parallelism}, one-forward-one-backward (1F1B) scheduling~\cite{pipedream,megatronsc21} needs many microbatches to fill its bubbles, whereas local reuse is largest when a group stays in few invocations.  

For \emph{memory}, prefix activations live as long as the invocation: suffix computation needs only the prefix \(K/V\) and its gradients, yet the remaining prefix activations have no interval in which they can leave HBM.  

Removing these conflicts requires moving reuse out of the invocation and into the training schedule, which a kernel change alone cannot do: it needs a prefix graph retained apart from any suffix, \(gK/gV\) accumulated across short-lived suffix graphs, and a delayed prefix backward that is provably correct (Section~\ref{sec:design}).

\section{Design}
\label{sec:design}

\pf{} is organized around one design choice: make prefix reuse a property of the training schedule rather than an optimization embedded in a packed graph, kernel, or traversal.  Section~\ref{sec:three-phase-schedule} presents the three-phase schedule and its savings over the baseline, Section~\ref{sec:design-summary} explains which drawbacks of prior designs the schedule removes and which challenges, C1--C3, it raises in their place, and Section~\ref{sec:training-time-kv-boundary} proves that the schedule is equivalent to standard full-sequence training.  

\subsection{Overall Three-Phase Schedule}
\label{sec:three-phase-schedule}

Our \emph{baseline} is a standard trainer without prefix reuse: it trains each trajectory \([P\Vert S^{(i)}]\) as an independent full sequence.  Within each sequence, the forward pass processes prefix and suffix tokens together at every layer, and the backward pass then traverses the layers in reverse, again processing both together.  To compare this computation with \pf{}, we regroup it into four parts that run in order, each spanning all layers: \emph{prefix forward} produces the prefix activations, including each layer's \(K/V\); \emph{suffix forward} reads that \(K/V\) and produces the suffix activations; \emph{suffix backward} produces suffix-side gradients and the gradients \(gK/gV\) of the prefix \(K/V\); and \emph{prefix backward} turns \(gK/gV\) into prefix-side gradients.  Because the prefix never attends to the suffix, prefix forward needs no suffix input, and suffix backward needs nothing from prefix backward, to which it hands only \(gK/gV\) (Section~\ref{sec:training-time-kv-boundary}).  The regrouping is therefore equivalent to the original forward and backward passes: it changes only the order of computation, not its amount, and the baseline's work for a rollout group is \(N\) copies of the four parts.

\begin{figure*}[t]
    \centering
    \includegraphics[width=.80\textwidth]{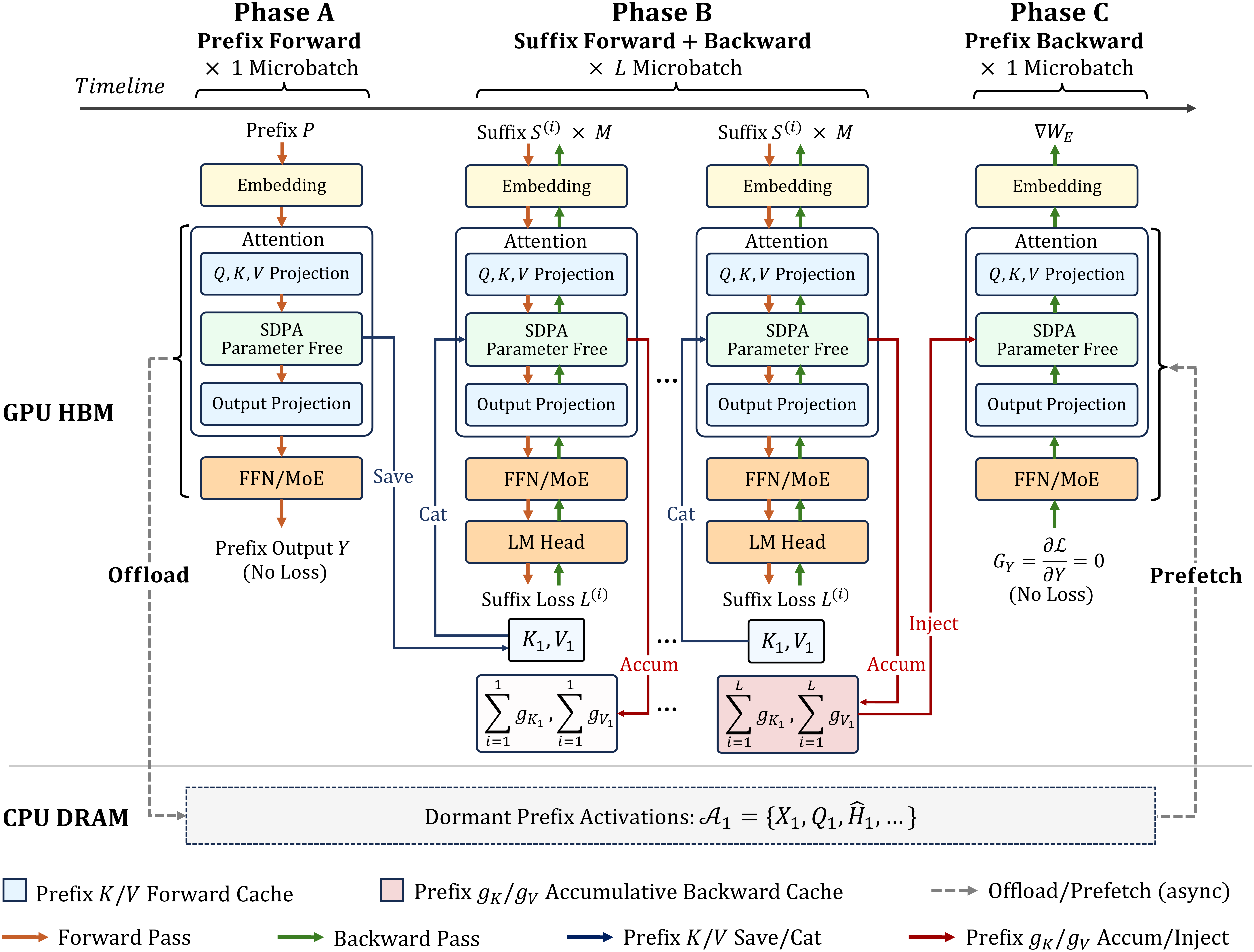}
    \caption{Overall \pf{} architecture and three-phase schedule: Phase A captures shared-prefix KV, Phase B runs suffix microbatches while accumulating \(gK/gV\), and Phase C runs one delayed prefix backward; dormant prefix activations may be offloaded during Phase B.}
    \Description{Overall PrefixFlow architecture showing prefix forward, cached prefix key/value tensors, suffix microbatch forward and backward passes, gradient key/value accumulation, optional offload, and final shared prefix backward.}
    \label{fig:three-phase}
\end{figure*}

We observe that the two prefix parts need not be repeated \(N\) times.  All \(N\) prefix forwards take the same input and produce the same result, so one suffices.  The \(N\) prefix backwards are not identical, since each receives a different \(gK/gV\); however, prefix backward is linear in \(gK/gV\), so the sum of their results equals a single prefix backward on the accumulated \(gK/gV\) (Section~\ref{sec:training-time-kv-boundary}).  \pf{} therefore keeps the order of the four parts but runs prefix forward and prefix backward once per rollout group, yielding the three phases below (Figure~\ref{fig:three-phase}).

\textbf{Phase A: prefix forward.}  \pf{} runs a normal forward pass over the shared prefix once.  Each transformer layer captures live prefix KV tensors and retains the prefix autograd graph for the delayed prefix backward.

\textbf{Phase B: suffix microbatches.}  \pf{} executes suffix work as ordinary training microbatches.  Each suffix forward reads the cached prefix KV, uses masks and position IDs equivalent to the baseline sequence \([P\Vert S^{(i)}]\), and then runs suffix backward.  The suffix graph produces ordinary suffix-side gradients and accumulates its prefix-side coupling gradients into \(gK/gV\) buffers.  The suffixes may be processed as padded batches, suffix-only packed segments, or separate microbatches; the prefix schedule is independent of this organization.

\textbf{Phase C: prefix backward.}  After all suffix microbatches finish, \pf{} injects the accumulated \(gK/gV\) cache into the retained prefix graph and runs one shared prefix backward.  Prefix-side and suffix-side gradients then contribute to the same logical optimizer step.

\textbf{Savings over the baseline.}  Per rollout group, the baseline's work is \(N\) copies of all four parts, whereas \pf{} performs the two suffix parts \(N\) times but each prefix part only once, plus one element-wise \(gK/gV\) accumulation per suffix.  This eliminates \(N-1\) prefix forwards and \(N-1\) prefix backwards and leaves the suffix work unchanged.  Caching by itself removes no computation; the schedule does: one \(K/V\) cache serves all \(N\) suffix forwards, and accumulated \(gK/gV\) let one prefix backward serve all \(N\) suffix backwards.  The price is state that must stay alive until Phase C: the per-layer prefix \(K/V\), read by every suffix microbatch; the remaining prefix saved tensors, consumed only in Phase C; and the \(gK/gV\) accumulators.  All are created in Phase A or B and freed after Phase C, while parameter gradients accumulate toward the optimizer step as usual, so no state is reused across rollout groups or weight updates.

The schedule also gives a compact speedup model.  Normalize the cost of one baseline full-sequence forward/backward to \(1\).  Let \(p\) be the one-time prefix forward/backward cost, \(s\) be the per-suffix Phase-B cost after prefix KV reuse, and \(o\) be the group-level overhead from scheduling, communication, offload, and compiler boundaries, all in the same normalized units.  For \(N\) suffix trajectories the baseline costs \(N\) and \pf{} costs \(p+Ns+o\), so
\[
    \mathrm{Speedup}(N)=\frac{N}{p+Ns+o}
        =\frac{1}{p/N+s+o/N}.
\]
The prefix term \(p\) and overhead term \(o\) are amortized by the trajectory count \(N\), while the suffix term \(s\) remains on every trajectory.  Therefore the overhead-free asymptotic upper bound is \(1/s\), and \pf{} is most beneficial when the reusable prefix fraction is large and the rollout group has enough trajectories to amortize the one-time work.

\subsection{Benefits and Challenges of Decoupling}
\label{sec:design-summary}
\label{sec:challenges}

\textbf{What decoupling solves.}  The three-phase schedule makes the shared prefix its own part of the training schedule: Phases A and C contain only prefix work, and Phase B trains the suffixes as ordinary microbatches.  The conflicts of Section~\ref{sec:why-schedule} then go away.  First, reuse no longer depends on how suffixes are organized: they may be padded, packed, or split across any number of microbatches, and a prompt's responses need not share one invocation.  Second, each phase holds a single computation mode, so CP balances each phase on its own and no suffix-heavy rank is left on the critical path (\S\ref{sec:impl-cp-kv-prefetch}).  Third, reuse no longer depends on how many microbatches 1F1B uses, so PP keeps as many as it needs (\S\ref{sec:impl-pp-stage}).  Fourth, prefix activations gain an interval to leave HBM, since Phase B reads and writes only the prefix \(K/V\) and \(gK/gV\) (\S\ref{sec:impl-phase-b-hbm}).  Separation alone, however, does not keep training correct, bound the HBM that prefix state occupies, or keep parallelism and MoE routing working; these are challenges C1--C3, one per axis of Table~\ref{tab:related-comparison}.

\textbf{C1: correctness without a shared graph.}  In one graph, backpropagation carries every suffix's gradient into the prefix automatically; across decoupled graphs, suffix gradients stop at the cached \(K/V\), and one prefix backward runs after all suffixes.  Correctness then hinges on two questions: whether anything besides \(K/V\) and \(gK/gV\) couples a suffix graph to the prefix graph, and whether one delayed prefix backward on the accumulated \(gK/gV\) equals the baseline's \(N\) prefix backwards under any batching or microbatch split.  The \emph{training-time KV boundary} answers both: nothing else crosses it, and prefix backward is linear in \(gK/gV\) (Proposition~\ref{prop:superposition}, Section~\ref{sec:training-time-kv-boundary}); hooks carry \(gK/gV\) into the retained prefix graph (\S\ref{sec:impl-kv-hook}).

\textbf{C2: lifecycle of dormant prefix state.}  Decoupling keeps prefix state alive from Phase A to Phase C.  During Phase B, the prefix \(K/V\) and \(gK/gV\) are \emph{hot}, read and written by every suffix microbatch, whereas the remaining prefix saved tensors are \emph{dormant}, needed only by the delayed prefix backward, yet they would occupy HBM for the whole phase.  Exploiting this interval raises three questions: which tensors can leave safely, how to move them out and back without stalling Phases A and C, and how to turn the released HBM into suffix budget.  \pf{} treats the KV boundary as a memory-lifecycle boundary whose completeness guarantees that no suffix touches a dormant tensor: it keeps the hot cache resident, offloads dormant tensors to a pinned host pool with asynchronous prefetch for Phase C, and reinvests the released HBM in suffix work (\S\ref{sec:impl-phase-b-hbm}).

\textbf{C3: composition with parallelism and MoE routing.}  The cache lives only while one rollout group is trained, under a fixed parallel configuration.  TP and EP need no change, because reuse occurs at the parameter-free scaled dot-product attention (SDPA) boundary (\S\ref{sec:impl-tp}, \S\ref{sec:moe-load-balance}); DP keeps each rollout group on one rank where possible and splits it, exchanging only \(K/V\) and \(gK/gV\), when that would unbalance ranks (\S\ref{sec:impl-dp-placement}).  CP, PP, and the MoE expert router, however, each break in a different way.  Under CP, every layer must all-gather the prefix \(K/V\) for suffix attention and reduce \(gK/gV\) back; issued on demand, this traffic exposes 43.77\,s of Phase-B communication, which prefetching the next layer's \(K/V\) cuts to 13.66\,s (Claim~1.2, \S\ref{sec:impl-cp-kv-prefetch}).  Under PP, global phase barriers drain and refill the pipeline, adding a fixed 8.2--9.2\,s bubble (Claim~1.3); \pf{} instead switches phases stage by stage, at the price of per-stage phase state (\S\ref{sec:impl-pp-stage}).  For MoE, Proposition~\ref{prop:superposition} needs a prefix token routed to the same experts in every trajectory, but \pf{} changes batch composition, so batch-dependent routing could diverge from the baseline; \pf{} fixes expert selection with Rollout Routing Replay, sharing one prefix mask per group and assuming dropless MoE layers (\S\ref{sec:moe-load-balance}).

\subsection{The Training-Time KV Boundary}
\label{sec:training-time-kv-boundary}

This section proves the two properties of the training-time KV boundary that Sections~\ref{sec:three-phase-schedule} and~\ref{sec:design-summary} rely on.  At this boundary, suffix microbatches read prefix \(K/V\) and return only prefix-side gradients \(gK/gV\) (the gradient-KV cache, abbreviated gKV) to the retained prefix graph.  The boundary is \emph{complete and compact}---nothing but \(K/V\) and \(gK/gV\) crosses it---so Phase B is independent of the prefix graph and the remaining prefix state is provably dormant.  It is \emph{linear}---for a fixed prefix forward trace, prefix backward is linear in the adjoints injected at it---so the single delayed prefix backward is both correct and cheap.

\smallskip
\noindent\textbf{Property 1: the boundary is complete and compact.}
For trajectory \(i\), split one layer's input as \(X^{(i)}=[X_1;X_2^{(i)}]\).  Under causal masking, suffix rows attend to prefix keys and values, while prefix rows do not attend to suffix tokens:
\[
    H_2^{(i)}
    =
    \mathrm{Attn}\!\left(
        Q_2^{(i)}, [K_1;K_2^{(i)}], [V_1;V_2^{(i)}]
    \right).
\]
Residual connections, normalization, and feed-forward network (FFN) layers are token-wise, so they add no prefix--suffix edge.  The same boundary holds in suffix backward: every suffix-row adjoint is computed from the suffix forward cache plus the cached \(K_1,V_1\), and no prefix-backward intermediate appears in any of these expressions.  In the reverse direction, attention backward returns exactly two prefix-side adjoints, where \(P_{21}^{(i)}\) and \(S_{21}^{(i)}\) are the suffix-to-prefix attention probability and score blocks and \(d\hat{H}_2^{(i)}\) is the suffix upstream adjoint:
\[
    \begin{aligned}
    gV_1^{(i)} &= \left(P_{21}^{(i)}\right)^\top d\hat{H}_2^{(i)},\quad
    gK_1^{(i)} &= \frac{1}{\sqrt{d_h}}\left(dS_{21}^{(i)}\right)^\top Q_2^{(i)} ,
    \end{aligned}
\]
and there is no third channel through \(Q_1\), because prefix queries participate only in prefix rows.

\emph{Consequences for the schedule.}  First, Phase B is legal as an independent phase: every suffix microbatch can run to completion before any prefix backward starts.  Second, the cross-boundary state is small and enumerable: per layer it is exactly \(K_1,V_1\) (read) and \(gK_1,gV_1\) (written), of size \(O(|P|\,d)\), so the remaining prefix saved tensors \(\mathcal{A}_1\)---all non-KV activations retained for prefix backward---are provably untouched throughout Phase B.  This dormancy is what licenses the Phase-B activation offload of Section~\ref{sec:impl-phase-b-hbm} and the compact KV/gKV exchange under CP and DP (Sections~\ref{sec:impl-cp-kv-prefetch} and~\ref{sec:impl-dp-placement}).

\smallskip
\noindent\textbf{Property 2: the boundary is linear.}
For \(L=\sum_i\alpha^{(i)}L^{(i)}\) with loss weights \(\alpha^{(i)}\), define
\[
    U^{(i)} =
    \bigl(\alpha^{(i)}dY_1^{(i)},\alpha^{(i)}gK_1^{(i)},\alpha^{(i)}gV_1^{(i)}\bigr),
\]
where \(dY_1^{(i)}\) is trajectory \(i\)'s direct prefix-token adjoint---zero for the common suffix-only actor loss, in which case the prefix still receives gradient through \(gK_1/gV_1\).  For a fixed prefix forward trace, the prefix backward operator \(B_{\mathrm p}\) is a vector-Jacobian product and is therefore linear in these incoming adjoints: with the saved forward statistics fixed, the backward of every projection, normalization, activation, and softmax is linear in its upstream adjoint.  This yields the central correctness statement of \pf{}:

\begin{proposition}[Prefix-gradient superposition]
\label{prop:superposition}
Fix the prefix tokens, model parameters, attention masks, position IDs, and the prefix forward trace, including expert routing in MoE layers (Section~\ref{sec:moe-load-balance}).  Over real arithmetic, running prefix backward once on the accumulated incoming adjoints produces the same prefix-side gradients as running prefix backward separately for every suffix and summing the results:
\[
    \sum_{i=1}^{N} B_{\mathrm p}\!\left(U^{(i)}\right)
    =
    B_{\mathrm p}\!\left(\sum_{i=1}^{N} U^{(i)}\right).
\]
\end{proposition}

\emph{Consequences for the schedule.}  The first consequence is correctness: with Property 1, which lets Phase B finish before Phase C begins, the proposition makes the three-phase schedule equivalent to the full-sequence baseline over real arithmetic.  Parameter gradients keep the baseline's prefix-plus-suffix split, e.g., \(dW_K=X_1^\top dK_1+\sum_i\alpha^{(i)}X_2^{(i)\top}dK_2^{(i)}\): Phase B accumulates the suffix terms, and Phase C computes the prefix term once from a \(dK_1\) that already contains every \(gK_1^{(i)}\).  In floating point, accumulation order and reduction trees differ, so the implementation target is agreement within datatype tolerance rather than bitwise identity (Claim~1.1).  The second consequence is the compute saving: prefix backward is dominated by general matrix multiplication (GEMM), costing \(O(|P|\,d^2 + |P|^2 d)\) per layer, whereas accumulating \(gK_1^{(i)},gV_1^{(i)}\), which suffix backward already produces, is an element-wise addition of \(O(|P|\,d)\) per layer.  Linearity therefore replaces \(N\) GEMM-dominated prefix backwards with \(N\) lightweight accumulations plus one prefix backward, dividing the GEMM term by \(N\) (the amortized \(p/N\) term of Section~\ref{sec:three-phase-schedule}).

\section{Implementation}
\label{sec:implementation}

We implement \pf{} on top of TorchTitan~\cite{torchtitan}.  The core functionality is factored into an independent module, \texttt{prefix\_flow}, so it can be applied non-invasively to any model supported by the TorchTitan stack.  A user only needs to wrap an already initialized model with a single entry point, \texttt{apply\_prefixflow\_ac}, to enable prefix KV capture, gradient-KV accumulation, delayed prefix backward, and selective activation offload.  This section turns the schedule in Section~\ref{sec:design} into a TorchTitan execution path: graph decoupling, prefix activation lifecycle management, and parallel execution.

\subsection{KV Capture, Hook Bridge, and Gradient Cache}
\label{sec:impl-kv-hook}

During Phase A, each \pf{} attention wrapper observes the live prefix \(K_1,V_1\) tensors produced by the ordinary prefix forward graph, but it does not expose that live graph directly to Phase B.  Instead, it creates detached cached leaves for suffix attention and registers coupling hooks on the live \(K_1,V_1\).  Each Phase-B suffix wave, i.e., one suffix microbatch step, builds a short-lived suffix graph, calls backward without retaining the live prefix graph, and accumulates the resulting coupling gradients in the cached leaves as \(gK_1,gV_1\).  Phase C then backpropagates through the retained prefix output once; the \texttt{register\_hook} callbacks add the accumulated cached-leaf gradients to the live prefix \(K/V\) tensors before the prefix graph is traversed.

This hook-based bridge is the implementation mechanism behind the algebra in Section~\ref{sec:training-time-kv-boundary}.  The cached leaves are not parameters and are not updated by the optimizer; they are gradient accumulators that connect independent suffix graphs to one retained prefix graph.

Phase B attention uses suffix queries with keys and values formed by concatenating cached prefix \(K/V\) and current suffix \(K/V\).  The attention mask must satisfy three conditions: each suffix query sees all prefix tokens, suffix queries see only causal suffix positions, and suffixes from different trajectories do not attend to each other.  Position IDs must match the baseline sequence \([P\Vert S^{(i)}]\); in particular, rotary position embedding (RoPE)~\cite{roformer} positions for suffix tokens continue after the prefix length rather than restarting from zero.

Phase B is the main \texttt{torch.compile}~\cite{pytorch2} target; Phase A capture and the Phase C delayed backward rely on autograd and saved-tensor hooks and stay outside the compiled region (Section~\ref{sec:discussion}).

\subsection{Prefix Activation Offload and HBM Reuse}
\label{sec:impl-phase-b-hbm}

\noindent\textbf{Hot/dormant separation.}
Phase separation gives prefix saved tensors a long dormant interval: they are produced in Phase A, unused throughout suffix computation in Phase B, and consumed only in Phase C.  \pf{} uses the training-time KV cache to separate the Phase-B prefix state into a hot set and a dormant set:
\[
\begin{array}{ll}
    \text{hot set:} & K_1, V_1, gK_1, gV_1,\\
    \text{dormant set:} & \mathcal{A}_1 \text{ saved tensors for prefix backward}.
\end{array}
\]
The hot set remains on GPU.  The dormant set can be packed to CPU pinned memory with saved-tensor hooks and restored only when Phase C needs it.  This is different from generic checkpointing or ZeRO-Offload-style tiering~\cite{chen2016checkpoint,vdnn,capuchin,zerooffload,zero}: \pf{} exploits a semantic lifecycle boundary created by phase separation, because Phase B needs the compact KV/gKV cache but not the full prefix activation stack.

\smallskip
\noindent\textbf{Pinned pool and transfers.}
Each rank stages offloaded tensors in a pinned host pool sized at about \(1.5{\times}\) the measured Phase-A saved-tensor footprint.  Prefix activations are packed to the pool after production and prefetched asynchronously before the matching prefix-backward layer; when HBM headroom is short, prefetch falls back to demand-load, so correctness never depends on overlap.

\begin{figure}[tbp]
    \centering
    \includegraphics[width=.90\columnwidth]{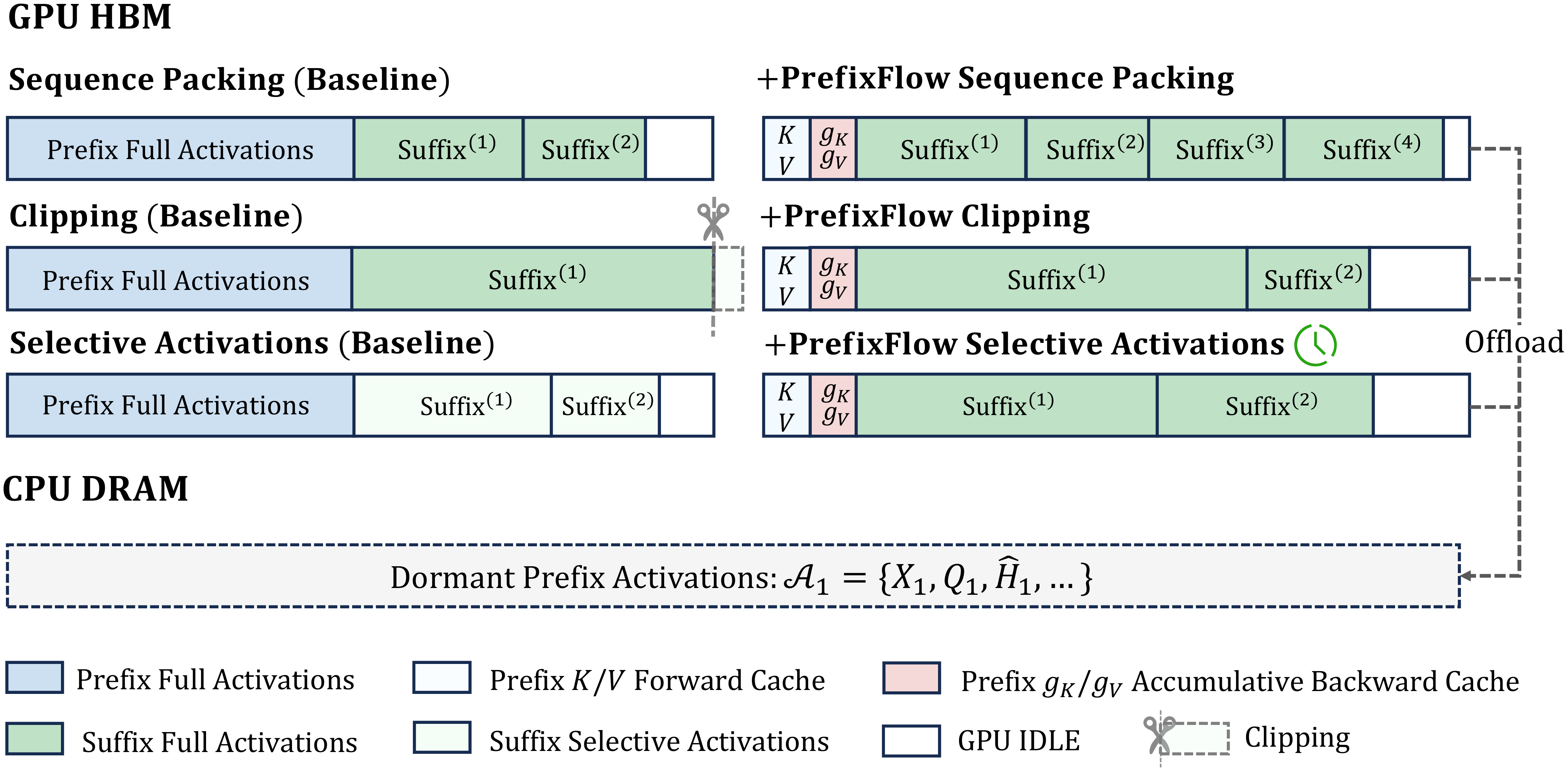}
    \caption{HBM reuse in Phase B.  By offloading dormant prefix activations, \pf{} keeps only the hot \(K/V\) and \(gK/gV\) cache resident and reinvests the released HBM into longer suffixes, more suffixes per wave, or more retained suffix activations.}
    \Description{Memory layout diagram showing that offloading dormant prefix activations leaves GPU memory for longer suffixes, more suffixes per Phase-B wave, or more suffix saved tensors.}
    \label{fig:hbm-packing}
\end{figure}

\smallskip
\noindent\textbf{Reinvesting the released HBM.}
The released HBM is useful because Phase B can reinvest it; Figure~\ref{fig:hbm-packing} illustrates the resulting Phase-B memory layout.  The first target is suffix length: if training is capacity-bound, the budget is spent on longer suffix tokens and their attention workspaces.  The second target is suffix wave size.  Suffixes can form one padded microbatch or a suffix-only packed sequence in which cached prefix \(K/V\) is a shared read-only block and the suffix-to-suffix part stays block-diagonal causal.  In both layouts, suffix backward accumulates into the same \(gK/gV\) buffers and Phase C runs prefix backward once.

The third target is fuller residency for expensive transient state.  The default selective activation checkpointing (SAC)~\cite{sac} policy recomputes cheap element-wise operations and saves only a memory-budgeted subset of expensive matrix-multiplication and attention-related tensors.  After prefix offload, Phase B can relax this policy: element-wise outputs may still be dropped, but expensive suffix activations are retained more aggressively when the budget allows.

\subsection{Tensor Parallelism}
\label{sec:impl-tp}

TP~\cite{megatronlm} partitions the model's projection and FFN weights and correspondingly shards hidden or head dimensions.  Prefix reuse operates at the SDPA operator: suffix attention reuses already materialized prefix \(K/V\) tensors, and SDPA itself has no trainable weights.  Thus TP only determines the local shard shape of the cached \(K/V\).  Each TP rank captures its own prefix \(K/V\) shard, concatenates the local prefix and suffix \(K/V\) along the sequence dimension, and otherwise follows the same \pf{} phase schedule.  No additional TP communication protocol is required beyond the model's existing sharded projections and output reductions.

\subsection{Expert Parallelism and MoE Routing}
\label{sec:moe-load-balance}

EP is orthogonal to prefix reuse for the same reason TP is: it partitions FFN experts~\cite{shazeer2017moe,gshard} and routes tokens to expert shards, but it never touches the attention \(K/V\) interface that carries reuse.  Prefix tokens execute their expert computation in Phases A and C under the existing EP routing, and suffix microbatches reuse prefix attention state in Phase B without any EP-specific cache protocol.

Expert selection is a separate question: Proposition~\ref{prop:superposition} needs one prefix forward trace to stand for all \(N\) trajectories, so a prefix token must select the same experts in every trajectory.  Recent MoE RL pipelines fix expert selection with Rollout Routing Replay~\cite{rolloutroutingreplay}: the inference engine records each token's top-\(k\) routing mask during rollout, and training replays the mask while computing gating weights from its own router logits, so the router still receives gradients.  With prefix caching, the engine stores prefix masks alongside the prefix KV cache, so all trajectories of a group share one prefix mask.  Prefix tokens then use the shared mask in Phases A and C and suffix tokens their own masks in Phase B, so expert selection depends neither on batch composition nor on the microbatch split, and the proposition applies under two assumptions: the prefix mask is shared within the group, and the MoE layers are dropless, unlike capacity-limited routing with token dropping~\cite{gshard,switchtransformer}, so no token's expert computation depends on other tokens.

\subsection{Context Parallelism}
\label{sec:impl-cp-kv-prefetch}

\begin{figure}[!t]
    \centering
    \includegraphics[width=.84\columnwidth]{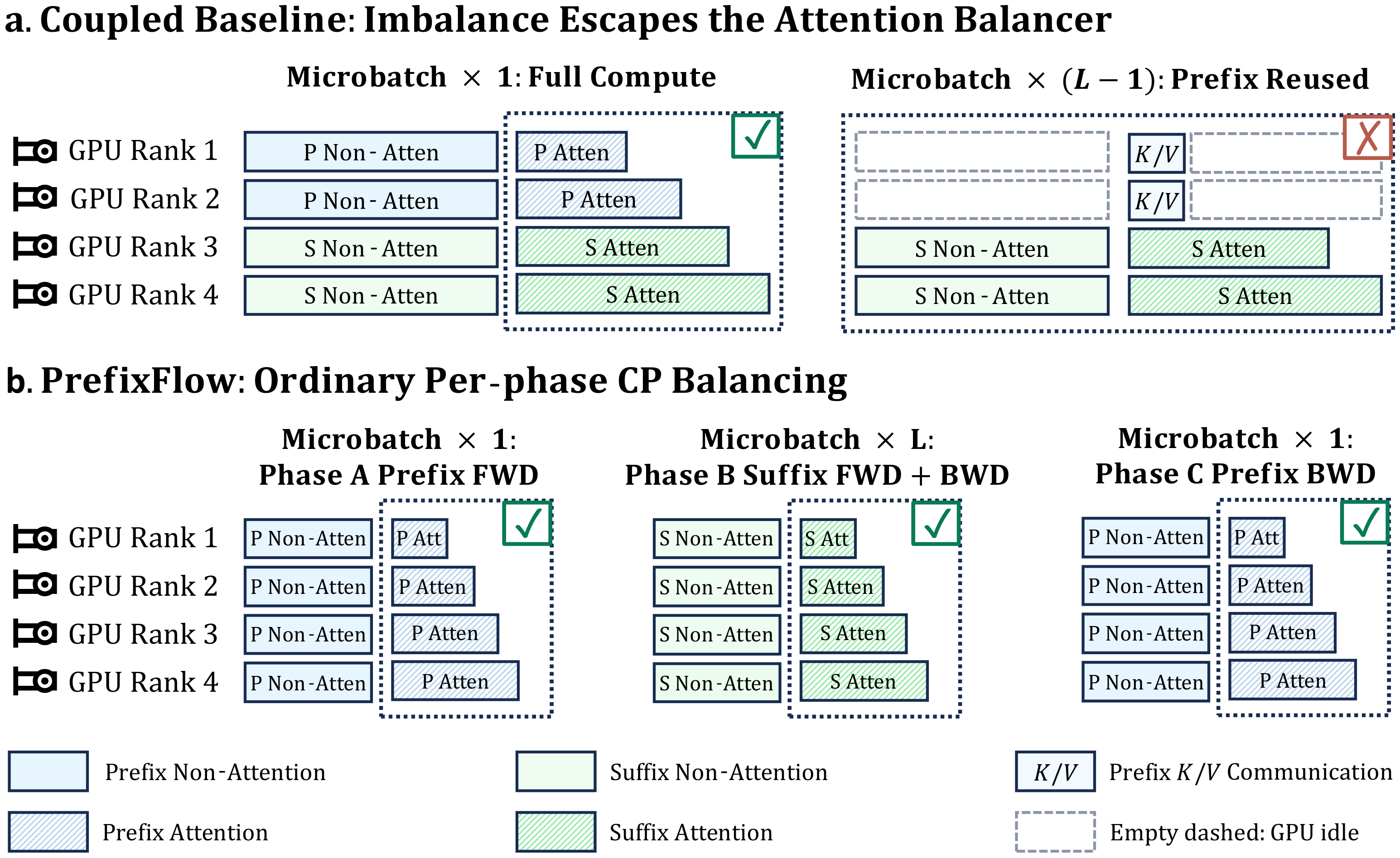}
    \caption{CP compatibility.  Phase separation lets existing CP balancing policies operate on prefix and suffix phases separately, while prefix \(K/V\) is treated as layer-local state for suffix attention.}
    \Description{Context-parallel diagram comparing a naive packed schedule with imbalanced prefix-owning ranks against a PrefixFlow phase-separated schedule.}
    \label{fig:cp-compat}
\end{figure}

CP partitions the sequence dimension for the attention/SDPA operator.  Even without prefix reuse, causal masking creates workload imbalance across sequence shards because later query blocks attend to longer histories.  Existing long-context CP systems such as DeepSpeed-Ulysses and RingAttention use all-to-all, ring, or blockwise schedules to balance this ordinary causal-attention workload~\cite{deepspeedulysses,ringattention}.  Coupled prefix-reuse schedules introduce a harder case (Section~\ref{sec:why-schedule}): after reuse, prefix tokens only provide their \(K/V\) while suffix tokens still execute the full per-token transformer flow, and a balancing policy that binds both kinds of tokens inside one traversal or packed graph cannot directly balance these two computation modes.  This imbalance is why reduced compute does not become a shorter step: the slowest suffix-heavy rank still sets the step time.

\pf{} avoids this coupled-balancing problem by separating prefix and suffix computation before CP balancing is applied.  In non-attention operators, Phase A and Phase C contain only prefix tokens and Phase B contains only suffix tokens, so token-wise work is balanced within a single computation mode.  In attention, causal imbalance still exists, but it is now the ordinary CP problem applied separately to prefix phases and suffix phases.  Phase B treats cached prefix \(K/V\) as temporary read-only state needed by suffix attention, not as active query tokens that define the suffix workload.  Therefore Ulysses-style and RingAttention-style balancing policies can be used independently on the prefix and suffix phases without changing either the CP algorithm or the \pf{} schedule.  Figure~\ref{fig:cp-compat} illustrates this phase-separated view.

CP also creates a compact communication requirement for the prefix KV itself.  Because CP shards the sequence dimension, each CP rank owns only a shard of the prefix \(K/V\), while suffix attention on every CP rank logically needs the complete prefix \(K/V\) to attend over the full shared prefix.  Before suffix attention consumes a layer, \pf{} materializes this full prefix-KV view by all-gathering the prefix \(K/V\) shards across the CP group.  Suffix attention then treats the gathered prefix block as read-only state and concatenates it with the local suffix \(K/V\).  During backward, the suffix-produced \(gK/gV\) for the gathered view is reduced back to the owning prefix shards before Phase C injects the accumulated gradients into the prefix graph.  This path exchanges only \(K/V\) and \(gK/gV\), not the dormant prefix activation stack.

The exchange is layer-local, so it can be scheduled like parameter prefetch.  While layer \(\ell\) computes suffix work, the runtime can issue the all-gather for layer \(\ell+1\)'s prefix \(K/V\) on a communication stream and release the gathered view after that layer's suffix backward has contributed its \(gK/gV\).  This overlaps the CP-induced KV transfer with useful layer computation instead of placing it entirely on the suffix critical path.

\subsection{Pipeline Parallelism}
\label{sec:impl-pp-stage}

\begin{figure}[tbp]
    \centering
    \includegraphics[width=.86\columnwidth]{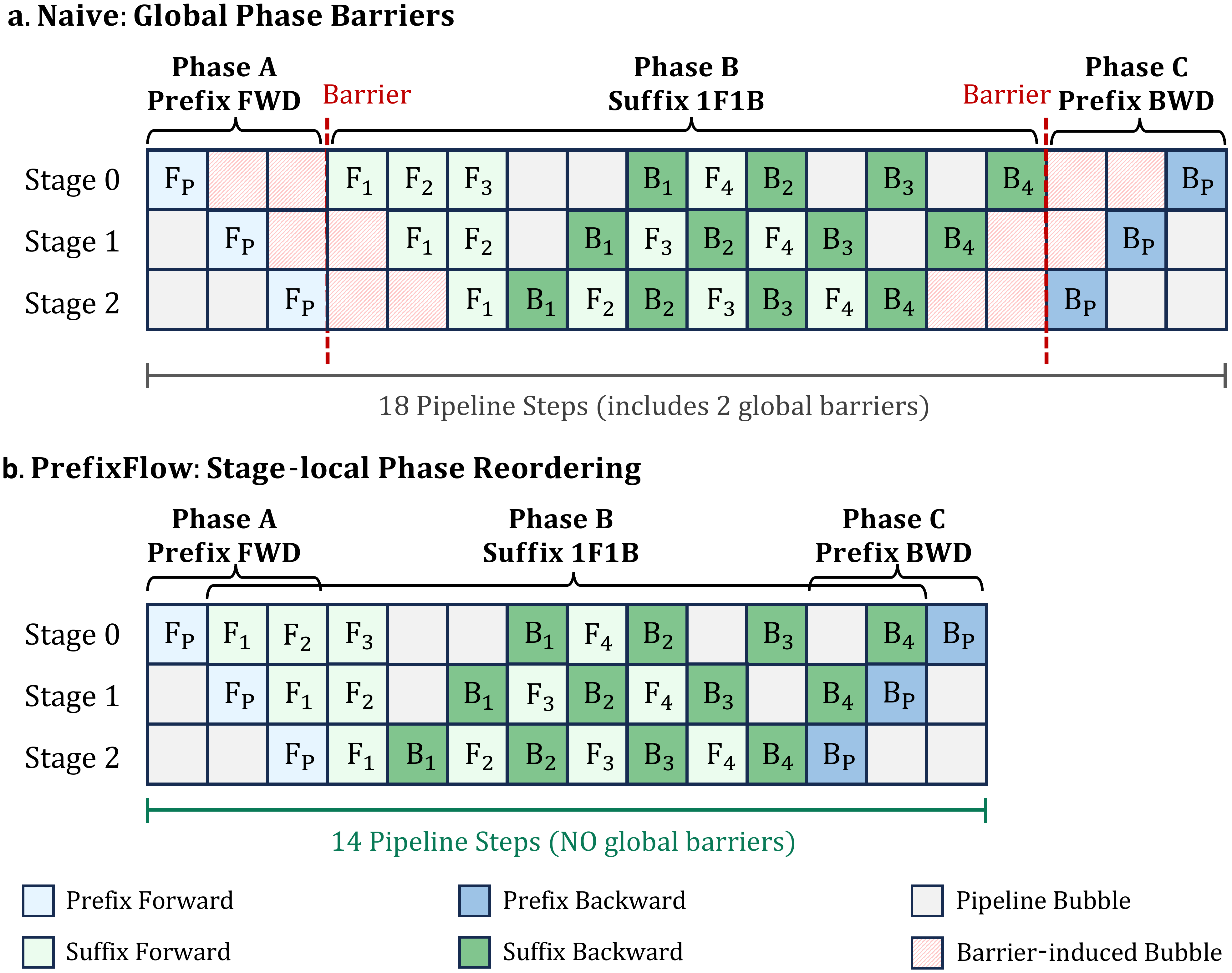}
    \caption{PP compatibility.  Stage-local switching schedules prefix work as pipeline work, avoiding a global Phase A--B--C barrier in tested PP schedules.}
    \Description{Pipeline parallelism diagram comparing global phase barriers with stage-local PrefixFlow switching.}
    \label{fig:pp-switch}
\end{figure}

Figure~\ref{fig:pp-switch} shows the PP scheduling issue.  A global Phase A--B--C barrier drains the pipeline before suffix work and refills it before prefix backward, creating bubbles around Phase B~\cite{gpipe}.  \pf{} instead switches each stage locally: prefix forward advances under ordinary send/recv dependencies, a stage enters the inner \(1\)F\(1\)B suffix schedule once its local prefix dependencies are satisfied, and prefix backward runs in reverse stage order.  Thus Phase A/C become local prefix work around ordinary Phase-B pipeline execution rather than global barriers.  Because every suffix microbatch reads the same cached \(K/V\), reuse does not depend on how many microbatches 1F1B uses, so the pipeline keeps enough microbatches to stay full.

\subsection{Data Parallel Placement}
\label{sec:impl-dp-placement}

Across DP ranks, \pf{} preferably keeps all trajectories for one prompt on one rank.  This maximizes reuse, avoids cross-rank cache traffic, and charges the shared prefix forward/backward once per rollout group rather than once per trajectory.  If this placement creates severe imbalance, the fallback splits the group and exchanges only compact prefix \(K/V\) plus the corresponding prefix-side \(gK/gV\) reduction; it never moves the full prefix activation stack.  FSDP~\cite{fsdp,zero} is orthogonal because parameter sharding does not change rollout-group placement.

\section{Evaluation}
\label{sec:evaluation}

We evaluate four claims.  \pf{} (1) preserves optimizer trajectories and schedules efficiently under representative parallel configurations, (2) accelerates a real RL loop with seed-level divergence, (3) speeds up the actor update in proportion to prefix reuse, and (4) converts dormant prefix activations into Phase-B HBM budget.  Claims 1.2, 1.3, 2.1, and 3 measure latency, Claim 4 measures HBM efficiency, and Claims 1.1, 2.1, and 2.2 test correctness.  Correctness follows from Proposition~\ref{prop:superposition} over real arithmetic; the experiments validate the BF16 implementation rather than prove the proposition by example.  Prefix reuse sets the ceiling of compute savings; the schedule cannot raise it, but it determines whether distributed execution reaches that ceiling and creates the memory opportunity.  Claims 1.2, 1.3, and 4 isolate this contribution of the schedule.

\subsection{Experimental Setup}

Experiments run on H20 clusters (96~GiB HBM per GPU) with up to four nodes and 32 GPUs, using a PyTorch 2.13 nightly build~\cite{pytorch} (\texttt{2.13.0.dev20260519}, CUDA 12.6 runtime, driver 535.247.01) and TorchTitan 0.1.0 at commit \texttt{43b4b94}.
We use Llama3-8B~\cite{llama3} for dense alignment, speed, and memory studies; Qwen3-32B for large-scale dense alignment; Qwen3-8B~\cite{qwen3} for end-to-end RL and trace replay; and Qwen3-MoE-30B-A3B for MoE/EP validation under Rollout Routing Replay (Section~\ref{sec:moe-load-balance}).  Unless stated otherwise, timing and memory studies use Llama3-8B with \(TP=2\) and \(CP=4\) on one node.  \textbf{The baseline performs a full forward/backward pass for every trajectory under the same loss, optimizer, masks, position IDs, precision, and parallel configuration.}  Correctness tables report loss, gradient, and after-update weight differences; timing uses rank-max GPU time averaged over five repeated runs; memory uses rank-max CUDA allocated HBM, which we refer to as peak HBM.

\subsection{Claim 1: PrefixFlow under Representative Parallel Configurations}
\label{sec:eval-claim1}

Claim~1 checks the distributed surfaces that can break reuse: trajectory equivalence, CP prefetch, and PP phase switching.

\subsubsection{Claim 1.1: Trajectory Alignment under Composed Parallelism}

\begin{table*}[t]
    \centering
    \caption{Claim~1.1 multi-step trajectory-alignment results.  Each step column reports max-absolute differences of gradients/losses/after-update weights (G/L/W) against the TorchTitan full-sequence baseline.  The MoE row runs under Rollout Routing Replay without an auxiliary load-balancing loss; both sides replay the same routing masks.}
    \label{tab:parallel-alignment}
    \scriptsize
    \setlength{\tabcolsep}{2pt}
    \begin{tabular*}{\textwidth}{@{\extracolsep{\fill}}l c c l l l c@{}}
    \toprule
    Model & Parallelism & Prefix+Suffix / \(N\) & \multicolumn{3}{c}{Max diff (G/L/W)} & Result \\
    \cmidrule(lr){4-6}
    & PP, FSDP, (E)TP, CP & & \multicolumn{1}{c}{Step 1} & \multicolumn{1}{c}{Step 10} & \multicolumn{1}{c}{Step 20} & \\
    \midrule
    Llama3-8B & (2,2,TP=2,2) & 1k+1k, \(N=4\) & 3.10e-05/1.91e-04/2.00e-06 & 3.72e-05/1.54e-04/1.41e-05 & 3.48e-05/1.14e-04/1.88e-05 & PASS \\
    Qwen3-8B & (2,2,TP=2,2) & 1k+1k, \(N=4\) & 2.30e-05/1.38e-04/2.00e-06 & 2.37e-05/2.82e-04/1.27e-05 & 2.07e-05/3.30e-04/1.73e-05 & PASS \\
    Llama3-8B & (1,1,TP=2,4) & 6k+6k, \(N=4\) & 1.31e-05/2.04e-05/1.99e-06 & 1.12e-05/1.10e-04/1.52e-05 & 1.38e-05/5.93e-05/2.57e-05 & PASS \\
    Qwen3-32B & (2,1,TP=8,2) & 2k+2k, \(N=8\) & 1.72e-05/1.63e-05/2.00e-06 & 1.72e-05/3.22e-04/1.90e-05 & 1.91e-05/6.70e-04/3.37e-05 & PASS \\
    Qwen3-MoE-30B-A3B & (2,2,ETP=2,2) & 3k+1k, \(N=8\) & 1.22e-04/1.39e-04/1.99e-06 & 1.18e-04/1.03e-04/1.60e-05 & 1.29e-04/1.46e-04/2.72e-05 & PASS \\
    \bottomrule
    \end{tabular*}
\end{table*}

Table~\ref{tab:parallel-alignment} reports five alignment runs.  Each run uses the same initialization, replay batch, loss, optimizer, and random seed as the TorchTitan full-sequence baseline.  At steps 1, 10, and 20, we compare loss, gradients before the optimizer step, and after-update weights.

The pass threshold is chosen from the executed precision path, not the measured error.  For matched elements \(x_i,y_i\), we use \(|x_i-y_i| \le \tau_{\mathrm{abs}} + \tau_{\mathrm{rel}}\max(|x_i|, |y_i|).\) BF16 mixed-precision updates include BF16 kernels, casts, and non-associative distributed reductions even when gradients accumulate in FP32, so we use \(\tau_{\mathrm{abs}}=10^{-3}\) and \(\tau_{\mathrm{rel}}=10^{-2}\).  These runs do not prove Proposition~\ref{prop:superposition} by example; they check that the finite-precision implementation stays within this tolerance.

All rows pass.  The dense rows cover two 8B architectures, two-node \(PP=FSDP=TP=CP=2\), a longer 6k+6k \(TP=2,CP=4\) setting, and Qwen3-32B under \(PP=2\), \(TP=8\), and \(CP=2\) on four nodes.  The MoE row runs Qwen3-MoE-30B-A3B under \(PP=FSDP=CP=2\) and expert tensor parallelism \(ETP=2\) with Rollout Routing Replay and no auxiliary loss: the baseline and \pf{} replay the same routing masks, so expert selection depends neither on batch composition nor on the microbatch split, and every difference stays below \(1.5{\times}10^{-4}\) through step 20.

\subsubsection{Claim 1.2: CP Phase-B Prefetch and Balance}

We validate CP prefix-\(K/V\) prefetch on Llama3-8B with \(PP=1,TP=1,CP=16\), a 16k-token prefix, an 8k-token suffix, and \(N=16\).  Figure~\ref{fig:claim1-parallel-schedule-ablations}(a) decomposes Phase B into compute, overlapped communication, and exposed communication.  Prefetch reduces mean Phase-B time from \(331.94\)s to \(316.90\)s by cutting exposed communication from \(43.77\)s to \(13.66\)s.  Demand all-gathers, issued when a layer's gathered prefix-\(K/V\) view is not ready, fall from 1024 to 32, the first layer's gathers.  Rank balance is preserved: the Phase-B spread (rank-max minus rank-min total Phase-B time) remains below 10ms for both on-demand and prefetch runs.

\subsubsection{Claim 1.3: PP Stage-Local Phase Switching}

We isolate PP scheduling with a Llama3-8B ablation over two H20 nodes.  Both variants run the same 4k+4k \pf{} computation and differ only in whether phase transitions are global barriers or stage-local switches.  We sweep \(N\) under \(PP=8,TP=2\) and \(PP=16,TP=1\), with \(CP=1\).

\begin{figure}[t]
    \centering
    \begin{minipage}[t]{0.51\columnwidth}
        \centering
        \includegraphics[width=\linewidth]{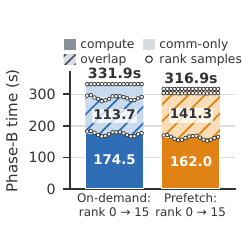}
        \par{\normalsize (a) CP prefetch}
    \end{minipage}\hfill
    \begin{minipage}[t]{0.47\columnwidth}
        \centering
        \includegraphics[width=\linewidth]{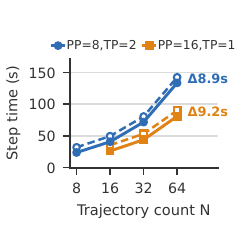}
        \par{\normalsize (b) PP stage-local switching}
    \end{minipage}
    \Description{Two side-by-side panels show Claim 1 parallel-scheduling ablations.  Panel a compares on-demand prefix key-value all-gather with prefetch under CP=16.  Panel b compares stage-local phase switching with a global-barrier schedule for PP=8, TP=2 and PP=16, TP=1; the annotated gaps stay around 8 to 9 seconds as trajectory count increases.}
    \caption{Claim~1 scheduling ablations.  (a) CP prefetch hides most prefix-\(K/V\) all-gather latency without rank imbalance.  (b) PP stage-local switching removes the fixed bubble from global Phase A--B--C barriers.}
    \label{fig:claim1-parallel-schedule-ablations}
\end{figure}

Figure~\ref{fig:claim1-parallel-schedule-ablations}(b) shows a nearly fixed gap between global-barrier and stage-local schedules: \(8.2\)--\(8.9\)s for \(PP=8,TP=2\) and about \(9.2\)s for \(PP=16,TP=1\).  As \(N\) increases, useful Phase-B work grows for both schedules, but the barrier penalty does not.  Stage-local switching removes this fixed drain/refill bubble while preserving ordinary pipeline dependencies.

\subsection{Claim 2: PrefixFlow Accelerates and Preserves a Real RL Loop}

Claim~2.1 integrates \pf{} into verl~\cite{verl} and runs 200 closed-loop GRPO steps; Claim~2.2 removes sampling feedback with a frozen-trace replay that isolates the trainer.

\subsubsection{Claim 2.1: End-to-End Closed-Loop RL}

We compare verl+TorchTitan against verl+\pf{} in a synchronous colocated GRPO loop---vLLM~\cite{vllm} rollout generation, policy evaluation, and the actor update---identical in every respect except the actor engine: Qwen3-8B with \(TP=2,CP=4\) on eight H20 GPUs (BF16, no compilation), trained for 200 steps on a long-prefix MuSiQue~\cite{musique} RAG dataset with two prompts per step, \(N=16\) rollouts per prompt, and a per-trajectory prefix ratio near \(95\%\).  Seeds and data order are pinned on both sides.  We chose MuSiQue, a public multi-hop RAG task, as a representative long-prefix workload; Claim~3 covers lower prefix ratios.

\begin{figure}[t]
    \centering
    \includegraphics[width=.82\columnwidth]{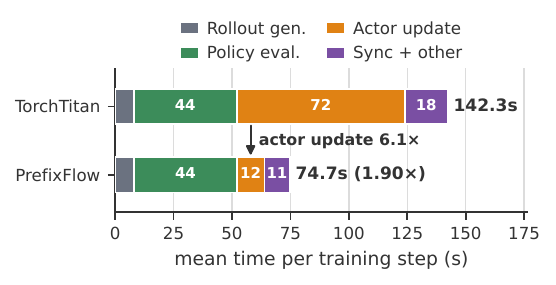}
    \Description{Horizontal stacked bar chart of mean time per training step.  The verl plus TorchTitan baseline takes 142.3 seconds: rollout generation, 44 seconds of policy evaluation, 72 seconds of actor update, and 18 seconds of synchronization and other work.  verl plus PrefixFlow takes 74.7 seconds, 1.90 times faster: rollout generation and policy evaluation unchanged, actor update reduced 6.1 times to 12 seconds, and 11 seconds of synchronization and other work.}
    \caption{Claim~2.1 end-to-end step-time breakdown over 200 closed-loop GRPO steps: \pf{} cuts the actor update \(6.1{\times}\) and the end-to-end step \(1.90{\times}\).}
    \label{fig:claim21-e2e-breakdown}
\end{figure}

\textbf{End-to-end performance.}  Figure~\ref{fig:claim21-e2e-breakdown} decomposes mean per-step time: \pf{} cuts the actor update \(6.1{\times}\) and the end-to-end step \(1.90{\times}\).  The gain lands entirely in the phase the schedule touches: rollout generation and policy evaluation are unchanged and act as a natural control group, while the small residual shrink in synchronization is consistent with shorter waits behind the faster actor update.  Two checks anchor the headline number.  First, the actor update is \(51\%\) of the baseline step, so eliminating it entirely would cap the end-to-end gain near \(2.0{\times}\); \(1.90{\times}\) is close to this Amdahl-style bound.  Second, the \(6.1{\times}\) actor-side reduction exceeds the largest sweep point of Claim~3 because this workload is more prefix-heavy (\(r\approx0.95\)) than any sweep split, consistent with the \(1/s\) bound of Section~\ref{sec:three-phase-schedule}; the scopes also differ, since Claim~3 times GPU forward/backward while the actor update here is wall-clock.

\begin{figure}[t]
    \centering
    \includegraphics[width=.96\columnwidth]{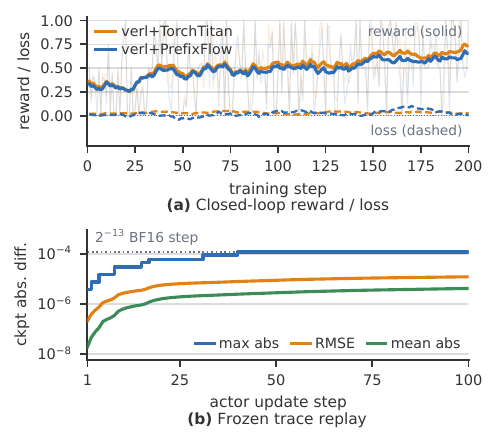}
    \Description{Two vertically stacked panels.  Panel a: single-axis chart over 200 training steps, with solid mean-training-reward curves of both systems rising from about 0.31 to about 0.65, forking locally but tracking the same trend, and dashed actor-loss curves oscillating around zero in the same band.  Panel b: log-scale chart of checkpoint absolute differences over 100 frozen replay steps, with the maximum absolute difference at a BF16-scale plateau of 1.22e-4, RMSE near 1.24e-5, and mean absolute difference near 4.24e-6.}
    \caption{Claim~2 training behavior and trainer alignment.  (a) Closed-loop mean training reward (solid) and actor loss (dashed) over 200 steps; thin lines are raw values, thick lines 21-step rolling means.  (b) Frozen trace replay against a verl/Megatron trace: across 399 checkpoint tensors and 8.19B actor parameters, the worst element stays at the BF16-scale plateau \(2^{-13}=1.2207{\times}10^{-4}\); root-mean-square error (RMSE) \(1.2433{\times}10^{-5}\) at step 100.}
    \label{fig:claim2-quality-replay}
\end{figure}

\textbf{Training quality.}  Figure~\ref{fig:claim2-quality-replay}(a) overlays the two runs' training reward and actor loss.  The curves fork, as they must: the prompts are identical step by step, but the sampled responses differ from the first update on.  What matters is that the fork stays bounded in behavior: both rewards climb from \({\approx}0.31\) to \({\approx}0.65\) with overlapping smoothed trajectories---window means \(0.522\) vs.\ \(0.487\), a gap of \(0.036\) on 32-rollout batches---and the losses oscillate in the same band.

\textbf{The fork is seed-level.}  The weight trajectories confirm that the fork is benign: on the identical seeded first batch the actor losses differ by \(3.2{\times}10^{-7}\), and after 200 free-running steps the worst checkpoint element still differs by only \({\sim}1.8{\times}10^{-4}\), the BF16-ulp level, while cross-system RMSE stays at \(1.37\)--\(1.38{\times}\) each run's own drift---matching the \(\sqrt{2}\) of two independent runs of the same training.  Replacing the trainer thus perturbs training by no more than a seed change; separating sampling randomness from trainer bias is the job of Claim~2.2.

\noindent\textbf{Claim 2.2: frozen trace replay isolates the trainer.}
The replay removes the rollout-side randomness identified above.  It uses the same MuSiQue GRPO recipe as Claim~2.1, with four configuration deltas---a verl+Megatron~\cite{megatronsc21} baseline, \(TP=8\), an 8{,}192-token prompt cap, and one prompt per step: verl+Megatron runs 100 Qwen3-8B actor updates, then exports trainer inputs and checkpoints, and \pf{} replays those frozen batches with the same actor loss and optimizer progression. The replay bounds trainer-side drift without claiming reward-level or raw-gradient identity.

Figure~\ref{fig:claim2-quality-replay}(b) shows no drift amplification over the replayed trajectory: the max curve moves in BF16 quantization steps, while at step 100 RMSE is \(9.8{\times}\) and mean absolute difference \(28.8{\times}\) below the worst element.  With sampling removed, the BF16-level tolerance of the closed loop thus holds under identical inputs.  Because the two experiments align against different production stacks (verl/TorchTitan and verl/Megatron), they also show that the \(K/V\)--\(gK/gV\) mechanism is not tied to one trainer implementation.

\subsection{Claim 3: Speedup Follows Prefix Reuse Opportunity}
\label{sec:eval-claim3}

\begin{figure}[t]
    \centering
    \includegraphics[width=.87\columnwidth]{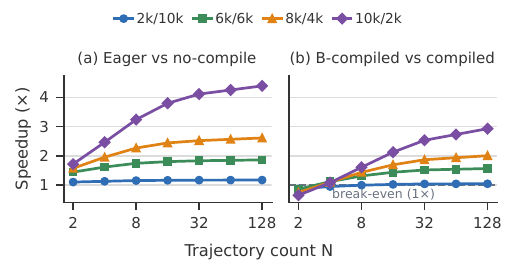}
    \Description{Two side-by-side line charts of PrefixFlow speedup versus trajectory count.  The left chart compares eager PrefixFlow to a no-compile baseline, and the right chart compares Phase-B-compiled PrefixFlow to a compiled baseline.  In both charts, speedup grows with trajectory count and with the prefix-heavy 10k/2k split.}
    \caption{Claim~3 speedup sweep.  (a) Eager \pf{} vs.\ no-compile baseline.  (b) Phase-B-compiled \pf{} vs.\ compiled baseline.}
    \label{fig:claim3-speedup-sweep}
\end{figure}

We sweep prefix ratio \(r=|P|/(|P|+|S|)\) and rollout group size \(N\) under the default configuration, holding total sequence length at 12,288 tokens.  These 12k-token sweeps are compute-bound, so Phase B feeds one trajectory at a time; larger suffix microbatches mainly increase activation pressure in this regime.  Speedup is baseline rank-max GPU forward/backward time divided by \pf{} rank-max GPU time.  Figure~\ref{fig:claim3-speedup-sweep} compares eager \pf{} against the no-compile baseline and Phase-B-compiled \pf{} against the compiled baseline; all \(N=2,4,8,16,32,64,128\) points are measured runs.

The eager panel, Figure~\ref{fig:claim3-speedup-sweep}(a), follows the reuse model: speedup increases with both \(N\) and prefix ratio.  The suffix-heavy 2k/10k split reaches only \(1.170{\times}\) at \(N=128\), while the prefix-heavy 10k/2k split reaches \(3.800{\times}\) at \(N=16\), \(4.115{\times}\) at \(N=32\), and \(4.395{\times}\) at \(N=128\).  Middle splits follow the same ordering.  The remaining gap to the overhead-free \(1/s\) bound corresponds to the amortized \(p/N\) and \(o/N\) terms of the speedup model in Section~\ref{sec:design}.

The compiled panel, Figure~\ref{fig:claim3-speedup-sweep}(b), is more conservative because the standard baseline benefits from whole-model compile, while the prototype compiles only Phase B and still pays an extra prefix/suffix concatenation copy.  Low-reuse \(N=2\) cases can be slower, but amortization overtakes these costs as \(N\) grows: at \(N=128\), compiled speedups are \(1.043{\times}\), \(1.565{\times}\), \(2.004{\times}\), and \(2.930{\times}\) from low to high prefix ratio.

\textbf{Where real workloads fall.}  The sweep spans prefix ratios of 17--83\%.  Chart QA and fixed-context RAG (75.8--87.6\%) fall near its upper end, and video RLVR, long-context RL/QA, and the Claim~2.1 workload lie above it (Table~\ref{tab:motivation-workloads}).  Math/code reasoning and SWE agents (2.3--12.8\%) lie below its low end, where even the 2k/10k split gains only \(1.170{\times}\) eager at \(N=128\).

\subsection{Claim 4: Training-Time KV Cache Converts Dormant Prefix State into Phase-B HBM Budget}

Claim~4 evaluates whether the prefix-activation offload of Section~\ref{sec:impl-phase-b-hbm} yields Phase-B headroom that suffix work can use.  We spend the released HBM in two ways: extending the sequence-capacity frontier (Claim~4.1) and accelerating suffix execution (Claim~4.2).

\begin{figure}[t]
    \centering
    \includegraphics[width=\columnwidth]{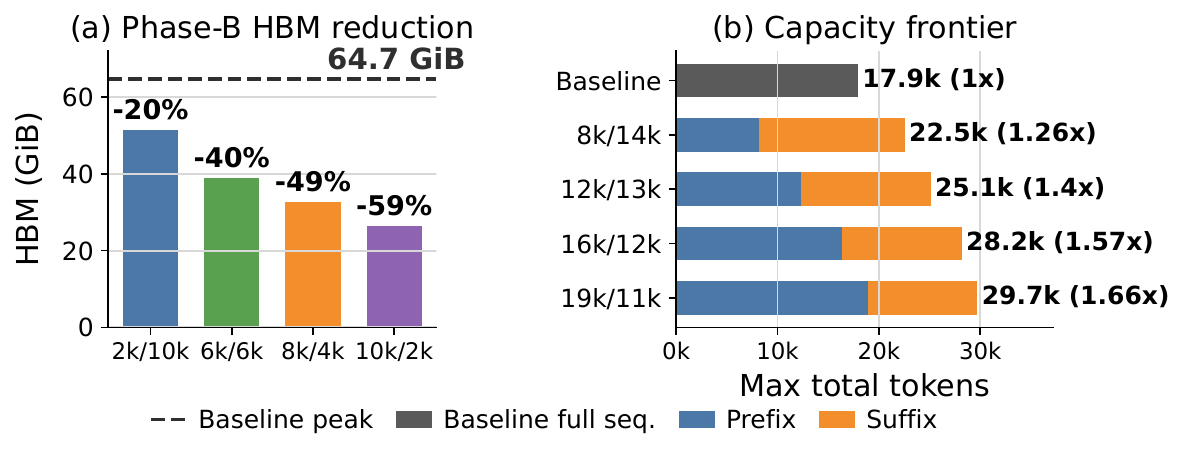}
    \Description{Two panels for Claim 4.1.  Panel (a) shows Phase-B allocated HBM falling from 51.57 GiB to 26.48 GiB as the prefix/suffix split changes from 2k/10k to 10k/2k, below a 64.75 GiB baseline peak.  Panel (b) shows the maximum total sequence length increasing from a 17.9k-token baseline to PrefixFlow frontiers from 22.5k to 29.7k tokens.}
    \caption{Claim~4.1 capacity headroom.  (a) At total length 12,288 and \(N=16\), \pf{} lowers Phase-B HBM as the reusable prefix grows.  (b) At \(N=2\), the released HBM expands the runnable frontier from 17,920 baseline tokens to 29,696 tokens.}
    \label{fig:claim41-hbm-capacity}
\end{figure}

\textbf{Claim 4.1: capacity headroom.}
Figure~\ref{fig:claim41-hbm-capacity} combines a fixed-total HBM diagnostic with a bracketed capacity-frontier search under the default configuration.

At fixed total length, \pf{} reduces Phase-B allocated HBM from the 64.7 GiB baseline peak to 51.6, 39.0, 32.8, and 26.5 GiB as the split moves from 2k/10k to 10k/2k, a \(20.3\%\) to \(59.1\%\) reduction (exact values in Figure~\ref{fig:claim41-hbm-capacity}(a)).  The frontier search shows that this allocator-level reduction becomes capacity: the default-SAC baseline fits 17,920 tokens, while \pf{} fits up to 29,696 tokens, \(1.657{\times}\) the baseline.  \textbf{At prefix length 18,944, the prefix alone exceeds the baseline's full-sequence limit, yet \pf{} still admits a 10,752-token suffix.}  The prefix itself can grow past that limit because offload and prefetch run asynchronously with computation: dormant prefix activations are packed out to the pinned pool while the prefix forward is still in flight, so the full prefix activation stack never co-resides in HBM.

\begin{figure}[t]
    \centering
    \includegraphics[width=\columnwidth]{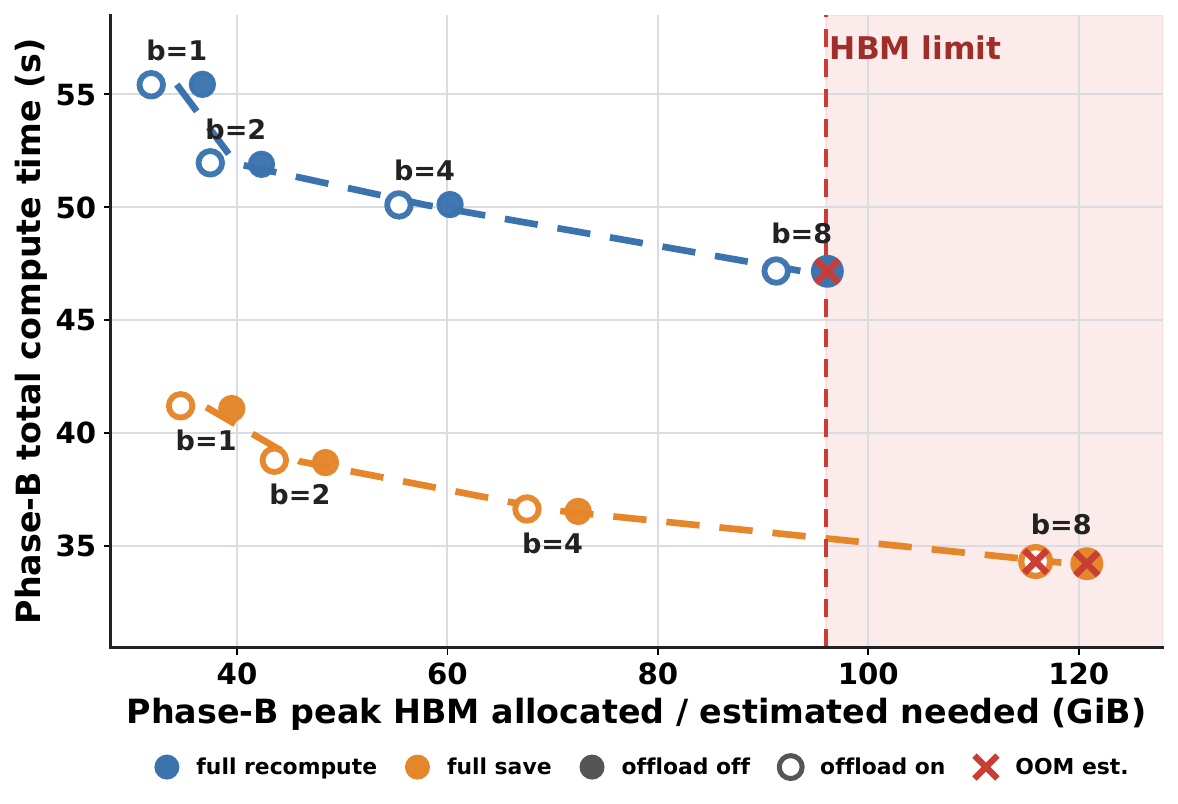}
    \Description{Tradeoff curves of Phase-B peak HBM versus Phase-B compute time, connected along suffix microbatch size, for full recompute and full save, with offload enabled or disabled.  Increasing suffix microbatch size moves points right and down.  A vertical HBM-limit line separates passing points from OOM estimates.  At b=8, full recompute without offload OOMs, but full recompute with offload fits; full save at b=8 OOMs even with offload.}
    \caption{Claim~4.2 suffix execution tradeoff.  Offload shifts points left by removing dormant prefix activations, while larger suffix waves and fuller suffix activation retention trade HBM for lower Phase-B time.}
    \label{fig:claim42-suffix-tradeoff}
\end{figure}

\textbf{Claim 4.2: faster suffix execution.}
Figure~\ref{fig:claim42-suffix-tradeoff} spends the same HBM budget on suffix execution with prefix 10,240, suffix 4,096, \(N=16\), no compilation, and a 64 GiB pinned pool, under the default configuration otherwise.  The suffix microbatch size \(b\) controls the number of Phase-B waves, and \emph{full save} retains suffix matrix-multiplication and linear outputs instead of recomputing them.

For \(b=1/2/4\), increasing \(b\) lowers Phase-B time and increases HBM; at fixed \(b\), full save cuts about 13.6--13.7s while spending additional HBM.  Prefix activation offload shifts successful points left by 4.86 GiB with little time change.  This shift matters at the feasibility boundary: without offload, both \(b=8\) policies run out of memory (OOM); with offload, full recompute at \(b=8\) fits at 91.25 GiB while full save still OOMs at an extrapolated 115.91 GiB.  Thus released HBM is a real allocation budget, usable for either longer sequences or lower suffix time.

\section{Discussion and Limitations}
\label{sec:discussion}

\textbf{Performance headroom.}  The prototype favors correctness and composability over peak performance: Phase A/C run eager because they rely on autograd hooks, and Phase-B attention pays one extra HBM copy to concatenate prefix and suffix \(K/V\).  Compile-friendly capture and a DualKV-style kernel~\cite{dualkv} that reads cached prefix \(K/V\) in place would remove both, so kernel-level and schedule-level reuse are complementary.  Nothing ties the design to TorchTitan; a Megatron port with fused kernels is the natural path to peak performance, and we avoid Megatron wall-clock comparisons, which would conflate the schedule with backend maturity.

\textbf{Scope and generalization.}  \pf{} is not a universal accelerator: suffix-dominated workloads gain little (Claim~3), and kernel-local reuse has lower overhead when a group fits one fused invocation and neither memory nor parallelism binds.  Beyond flat prefix sharing, each internal node of tree-structured RL data can carry its own \(K/V\) cache and \(gK/gV\) accumulator~\cite{arealdta,treetrainingv5}, and a long single sequence can train segment by segment, accumulating \(gK/gV\) into earlier segments as Proposition~\ref{prop:superposition} prescribes.  Rollout Routing Replay exploits the same property, caching prefix routing masks with the prefix KV cache~\cite{rolloutroutingreplay}, because state fixed by a prefix does not change across its continuations.

\section{Conclusion}

\pf{} turns repeated-prefix RL training into a schedule-level cache problem: one prefix forward and one delayed prefix backward serve an entire rollout group, and only \(K/V\) and \(gK/gV\) stay resident during suffix training.  TP and EP need no change, CP, PP, and DP need local rules with stated costs, and MoE models compose through Rollout Routing Replay; on a long-prefix RAG workload, end-to-end RL step time falls by \(1.90{\times}\).  The benefit is confined to workloads whose shared prefixes dominate compute; within that regime, training-time \(K/V\) plus \(gK/gV\) is a reusable systems boundary, and flat prefix reuse is its first application.

\begin{acks}
We thank the anonymous reviewers and our shepherd for their constructive suggestions.  We also thank Tencent Hunyuan and the TACC cluster~\cite{tacc} at The Hong Kong University of Science and Technology for providing the GPU resources used in our experiments.  Kai Chen and Guangming Sheng are the corresponding authors.
\end{acks}

\clearpage
\bibliographystyle{ACM-Reference-Format}
\bibliography{references}


\end{document}